\documentclass[sigconf,natbib=false,authorversion,nonacm]{acmart}
\usepackage{mathtools,amsthm,amsfonts,nicefrac}
\usepackage{csquotes}
\usepackage[backend=biber,style=ieee,sortlocale=en-US,url=true,doi=false,date=year,minnames=2,maxnames=6,isbn=false, sortcites, dashed=false]{biblatex}
\usepackage{tikz}
\usepackage[binary-units=true]{siunitx}
\usepackage{enumitem}
\usepackage{microtype}
\usepackage{booktabs}
\usepackage{hyperref}

\renewcommand{\baselinestretch}{0.96}

\newcommand{\ket}[1] {\ensuremath{| #1 \rangle}}
\newcommand*{\qop}[1]{\ensuremath{\mathit{#1}}}

\usetikzlibrary{backgrounds,decorations.pathmorphing,decorations.pathreplacing,positioning,matrix}

\hypersetup{
	pdftitle={Mixed-Dimensional Qudit State Preparation Using Edge-Weighted Decision Diagrams},
	pdfsubject={61st Design Automation Conference (DAC) 2024},
	pdfauthor={Kevin Mato, Stefan Hillmich, and Robert Wille} 
}

\newtheorem{example}{Example}

\let\origtheequation\theequation
\makeatletter
\def\tagform@#1{\maketag@@@{\ignorespaces#1\unskip\@@italiccorr}}
\makeatother
\renewcommand{\theequation}{(\origtheequation)}

\AtBeginDocument{%
}
\begin{document}

\title{Mixed-Dimensional Qudit State Preparation\\ Using Edge-Weighted Decision Diagrams}

\author[Kevin Mato, Stefan Hillmich, and Robert Wille]{Kevin Mato$^*$\hspace{3.0em}Stefan Hillmich$^{\ddag}\hspace{3.0em}$Robert Wille$^{*\ddag}$\hspace{3.0em}}
\affiliation{%
   \institution{\vspace{4pt}$^*$ Chair for Design Automation, Technical University of Munich, Germany}
   \country{}
}
\affiliation{%
  \institution{$^\ddag$ Competence Center Hagenberg (SCCH) GmbH, Austria}
  \country{}
}
\email{kevin.mato@tum.de, stefan.hillmich@scch.at, robert.wille@tum.de}
\email{https://www.cda.cit.tum.de/research/quantum/}

\begin{abstract}
    Quantum computers have the potential to solve important problems which are fundamentally intractable on a classical computer.
    The underlying physics of quantum computing platforms supports using multi-valued logic, which promises a boost in performance over the prevailing two-level logic.
    One key element to exploiting this potential is the capability to efficiently prepare quantum states for multi-valued, or \emph{qudit}, systems.
    Due to the time sensitivity of quantum computers, the circuits to prepare the required states have to be as short as possible.
    In this paper, we investigate quantum state preparation with a focus on mixed-dimensional systems, where the individual qudits may have different dimensionalities.
    The proposed approach automatically realizes quantum circuits constructing a corresponding \mbox{mixed-dimensional} quantum state. To this end, decision diagrams are used as a compact representation of the quantum state to be realized.
    We further incorporate the ability to approximate the quantum state to enable a finely controlled trade-off between accuracy, memory complexity, and number of operations in the circuit.
    Empirical evaluations demonstrate the effectiveness of the proposed approach in facilitating fast and scalable quantum state preparation, with performance directly linked to the size of the decision diagram.
    The implementation is freely available as part of \emph{Munich Quantum Toolkit}~(MQT) at \href{https://github.com/cda-tum/mqt-qudits}{\emph{github.com/cda-tum/mqt-qudits}}.
\end{abstract}

\maketitle
\section{Introduction}
\label{sec:introduction}
Quantum computing is capable of addressing problems intractable to classical computers, thanks to its distinct paradigm.
Examples include, Shor's algorithm~\cite{shor} to factorize integers, Grover's search~\cite{grover1996fast} for unstructured data, and evaluating possible materials in quantum chemistry~\cite{Cao_2019}. 
Worldwide, numerous academic and industrial research groups, including Google, IBM, and Microsoft, are actively exploring its potential and pushing the boundaries of its current capabilities. 
Until now, most applications have centered around \mbox{two-dimensional} qubit systems, which has led to the underutilization of the abundant potential offered by higher-dimensional systems present in various physical realizations of quantum computers. While constructing circuits with higher dimensions poses challenges, it also brings forth numerous benefits and serves as a more conducive platform for algorithms like quantum simulation.

The concept of higher-dimensional systems and its corresponding theory has existed for a considerable period~\cite{wang2020qudits}.
Fundamentally, qudits offer denser information storage and a broader range of operations than qubits. This has led to the successful demonstration of basic control in physical platforms such as trapped-ions~\cite{ringbauer2021universal}, to photonic systems~\cite{Hu2018a} and \mbox{superconducting circuits}~\mbox{\cite{Kononenko2020}}.

Recent developments in quantum algorithms have shown that multi-level logic is a more natural architecture for implementing complex applications~\cite{deller2022quantum}. 
Simulations of models representing fermion-boson interactions on mixed-dimensional quantum computers could enable real-time simulations of quantum electrodynamics and other field theories with continuous or larger symmetry groups~\cite{quditbasedQED, fermBoson}.
This is possible with the potentially reduced circuit complexity due to the temporary expansion of the Hilbert~space~\cite{shortcuts}. 

As the arrival of new and robust multi-level quantum devices and innovative quantum algorithms is anticipated, it is reasonable to expect encountering familiar challenges similar to those of qubit systems. Accordingly, methods, e.g., for the simulation~\cite{mato_sim} or compilation~\cite{Mato22, Mato23, mato2023compression}, of mixed-dimensional quantum circuits have been proposed recently. In this context, also methods for state preparation gain relevance especially for promising applications such as quantum simulation and quantum machine learning~\cite{kerenidis2016, Low2019, Childs2018}, that heavily rely on specific initial states to kickstart their processes effectively. Tackling the state preparation problem is crucial not only for facilitating quantum simulations but also for gaining insights into the behavior of specific states that have not yet been extensively studied in qudit systems, including aspects like entanglement. By addressing this obstacle, the full potential of various applications for quantum technologies can be unlocked.  

In this paper, we present a novel state preparation synthesis approach to compile mixed-dimensional quantum states with few operations and controls, later transpiled for an architecture-specific gate-set.
To this end, the following three contributions are made:
\begin{itemize}
    \item We explore the utilization of edge-weighted decision diagrams for a compact and efficient representation of \emph{mixed-dimensional} states.
    \item We propose an automated procedure to generate quantum circuits for the construction of arbitrary quantum states.  
    \item Further, we explore optimized circuit realizations, by leveraging the approximation of decision diagrams.
\end{itemize}

We demonstrate the advantages and potential of the proposed method by compiling a diverse range of quantum states, both uniform and random. 
These states were synthesized into quantum circuits with qudits of \mbox{mixed-dimensions}. 
For the first time, this showcases the automatic and efficient generation of \mbox{mixed-dimensional} quantum circuits that can construct arbitrary states. Additionally, by incorporating approximations, these circuits can be further optimized without significantly compromising their fidelity.

The remainder of this paper is organized as follows:
\autoref{sec:background} gives a brief background on quantum computations with qubits and qudits.
\autoref{sec:motivation} describes the considered problem, reviews the state of the art, and summarizes the contribution of this paper.
\autoref{sec:implementation} introduces the proposed approach for synthesizing quantum circuits for qubit-qudit circuits in detail.
\autoref{sec:evaluation} evaluates the proposed approach.
Finally, \autoref{sec:conclusion} concludes the paper.

\section{Background}
\label{sec:background}

In this study, we establish the groundwork for effectively preparing states in mixed-dimensional quantum systems. Initially, we offer a concise overview of quantum information, with a special emphasis on multi-level quantum logic.

In classical computations, information is encoded using \emph{bits}, binary digits that exist in either the 0 or 1 state. When we transition to quantum computing, we introduce the concept of a \emph{qubit} (quantum bit) as the fundamental unit of information. Unlike classical bits, qubits possess a remarkable feature--they can exist in a superposition, representing almost any linear combination of $\ket{0}$ and $\ket{1}$ (using Dirac's notation). This characteristic forms the core distinction between quantum and classical computing paradigms.

Qubits are typically engineered by confining the \mbox{multi-level} configuration of the underlying physical components that store information. Consequently, these systems inherently support \emph{\mbox{multi-level} logic}, where the basic information unit is referred to as a \emph{qudit} (quantum digit).
A qudit represents the quantum counterpart of a $d$-ary digit, where $d\geq 2$. Its state can be expressed as a vector within the $d$-dimensional Hilbert space $\mathcal{H}_d$, and it can be written as a linear combination \mbox{$\ket{\psi} = \alpha_0 \cdot \ket{0} + \alpha_1 \cdot \ket{1} + \ldots + \alpha_{d-1} \cdot \ket{d-1}$}, or simplified as vector $ \ket{\psi} = \begin{bsmallmatrix} \alpha_0 & \alpha_1 & \ldots & \alpha_{d-1}\end{bsmallmatrix}^\mathrm{T} $, where $\alpha_i \in \mathbb{C}$ are the amplitudes relative to the orthonormal basis of the Hilbert~space---given by the vectors $\ket{0}, \ket{1},\ket{2},\ldots, \ket{d-1}$.
The squared magnitude of an amplitude $|\alpha_i|^2 $ defines the probability with which the corresponding basis state $i$ will be observed when measuring the qudit. 
Since the probabilities have to add up to $1$, the amplitudes have to satisfy $\sum_{i=0}^{d-1} |\alpha_i|^2 = 1$.

\begin{example}\label{ex:state}
    Consider a system of one qudit with only three energy levels (also referred to as \emph{qutrit}).
    The quantum state $\ket{\psi} = \sqrt{\nicefrac{1}{3}}\cdot\ket{0} + \sqrt{\nicefrac{1}{3}}\cdot\ket{1} + \sqrt{\nicefrac{1}{3}}\cdot\ket{2}$ is a valid state with equal probability of measuring each basis. 
    Equivalently, the quantum state may be represented as vector $\sqrt{\nicefrac{1}{3}}\cdot \begin{bsmallmatrix} 1 & 1 & 1\end{bsmallmatrix}^\mathrm{T}$.
\end{example}

Quantum computing stands apart from classical computing due to two critical characteristics: superposition and entanglement. When we refer to a qudit being in a \emph{superposition} of states within a specific basis, it means that at least two amplitudes, relative to this basis, are non-zero. In simpler terms, the qudit can exist in multiple states simultaneously. On the other hand, \emph{entanglement} represents a unique kind of superposition that arises from interactions in multi-qudit systems. Entanglement entails a robust form of quantum correlation, where the entire system's state carries encoded information, and it becomes impossible to extract information from individual qudits separately.
The state of a single $d$-level qudit system can be manipulated by operations which are represented in terms of \mbox{$d \times d$-dimensional} unitary matrices $U$, i.e.,~matrices that satisfy $U^\dagger U = U U^\dagger = I$.
The state after the application of $U$ can be determined by multiplying the corresponding input state from the left with the matrix~$U$.

\begin{example}\label{ex:hadamard}
     Consider a three-level qudit (i.e.,~a qutrit) initially in the state $\ket{0}$. Applying the Hadamard operation~$H$ to it yields the output state shown before in \autoref{ex:state}, i.e.,
    \begin{align*}H\cdot \ket{0}=
        \frac{1}{\sqrt 3}
        \begin{bmatrix}
            1 & 1 & 1 \\
            1 & e^{\frac{2\pi}{3}} & e^{\frac{-2\pi}{3}} \\
            1 & e^{\frac{-2\pi}{3}} & e^{\frac{2\pi}{3}}
        \end{bmatrix} \cdot
        \begin{bmatrix}
            1 \\
            0 \\
            0 
        \end{bmatrix} 
        = \frac{1}{\sqrt 3}
        \begin{bmatrix}
            1 \\
            1 \\
            1 
        \end{bmatrix}.
    \end{align*}
\end{example}

\section{Motivation}
\label{sec:motivation}
This section revisits the problem of state preparation, particularly in mixed-dimensional quantum circuits. Then, we review relevant prior research and highlight the contribution of this work, which focuses on providing a scalable state preparation method by leveraging mixed-dimensional decision diagrams.

\subsection{Considered Problem}
\begin{figure}[tpb]
    \centering
    \includegraphics[width=0.85\linewidth]{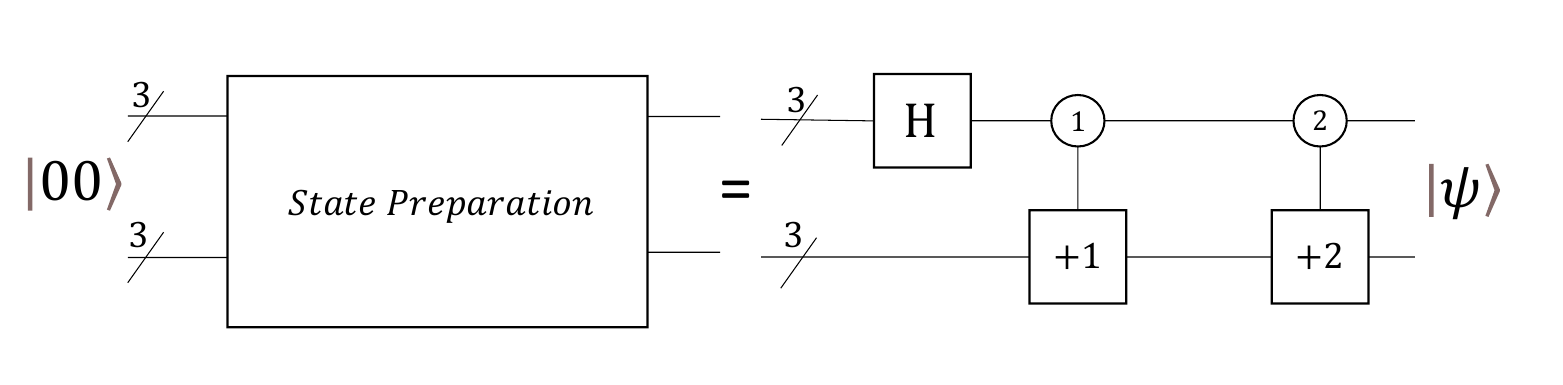}
    \caption{The state preparation algorithm for constructing $\ket{\psi}$, compiled into a sequence of single and two-qudit operations, in this case a two-qutrit GHZ state.}
    \label{fig:problem}
    \vspace{-1em}
\end{figure}

Quantum state preparation is a topic of great interest, both theoretically and practically. It plays a crucial role in determining the efficiency of inputting classical data into a quantum computer. Additionally, it serves as a critical subroutine for various quantum algorithms, including \mbox{well-known} ones~\mbox{\cite{qpe, grover1996fast, shor}}, as well as quantum machine learning~\mbox{\cite{Harrow2009, Lloyd2014, kerenidis2016}}, and Hamiltonian simulations~\mbox{\cite{Childs2018, Low2019}}. Formally, the task of quantum state preparation involves preparing a given quantum state $\ket{\psi}$ from an initial product state $\ket{0}^{\otimes N}$, using single and two-qudit gates of different types, as shown in \autoref{fig:problem}. However, quantum operations are prone to errors due to factors such as limited qudit connectivity, decoherence, and gate infidelity. 
These factors lead to inaccuracies in quantum computations, necessitating the development of methods and techniques that can achieve reliable results by minimizing the number of operations. 

Hence, the state preparation problem is to find the shortest sequence of quantum operations that can implement a state~$\ket{\psi}$, starting from the product state $\ket{0}^{\otimes N}$, while achieving a desired quantum state fidelity, in the most computationally efficient fashion.

\begin{example}
    Let's consider the realization of the quantum state $\frac{1}{\sqrt{3}}(\ket{00} + \ket{11} + \ket{22})$, i.e.,~a GHZ state, in a qutrit-qutrit system. The state is entangled and a superposition of three basis states $\ket{00},\ket{11},\ket{22}$. The problem is to find a sequence of operations that can transform the state $\ket{00}$ into this state. \autoref{fig:problem} shows a possible result: the qutrit Hadamard generates the superposition as shown in \autoref{ex:hadamard}. Then, controlled operations on the state of the first qutrit manipulate the second one. Each operation is controlled by the level inside the circle and imposes an increment based by \enquote{+1} or \enquote{+2}.
\end{example}

\subsection{Related Work}
While previous methods have been proposed for preparing arbitrary quantum states, their focus has been on qubit quantum circuits and tailored to use cases or particular state classes~\mbox{\cite{Niemann2016LogicSF, Moza2022, Mozafari2022}}. However, recent advancements in circuit depth and complexity have brought state preparation qubit circuits to a promising stage, albeit at the expense of an exponential number of ancillary qubits \cite{optZhang2022}.
In the context of state preparation, decision diagrams have been utilized with success for specific classes of states, such as uniform states, cyclic states, and arbitrary states, although with the need for ancilla qubits~\cite{Moza2022, Mozafari2022}. These studies demonstrate the effectiveness of decision diagrams in automating state preparation. Additionally, research has been conducted on searching for physical experiments designed to construct states directly~\cite{Krenn2016}.
While scalable algorithms have been presented for the case of prime-dimensional qudits \mbox{w-states}, or their qubit-embedded version~\cite{yeh2023scaling}, the investigation and development of scalable automated procedures, particularly optimized ones, for qudit circuits and mixed-dimensional systems, remain unexplored.

\subsection{Contribution}
In this paper, we investigate state preparation methods for mixed-dimensional quantum systems, facilitating various quantum algorithms on such architectures while automating complex routines involved in quantum computations. 
The proposed approach aims to enhance circuit efficiency in a scalable manner.
To achieve this, we introduce a tool that enables automatic and efficient quantum state preparation using decision diagrams for mixed-dimensional systems.

To realize such method, we make the following three contributions:
\begin{itemize}

\item The first contribution of this work is the exploration of the problem of state preparation by leveraging a recently investigated data structure, called edge-weighted decision diagrams with a variable number of successors. This data structure enables a more accurate and realistic representation of diverse quantum systems, accommodating qubits and qudits of different dimensions, akin to the natural diversity observed in complex quantum simulations. This is significantly different to traditional decision diagrams~\cite{Miller2006}, which were restricted to handling uniform-sized information units. Moreover, this study holds particular significance as there is no prior work on state preparation based on decision diagrams with a variable number of successors.

\item The second contribution of this work focuses on the synthesis of a high-level quantum circuits, which facilitates the automated state preparation procedure in an efficient manner. The complexity of this routine is linear in the number of nodes of the decision diagram. This significant advancement plays a crucial role in constructing states that would otherwise be exceedingly difficult and time-consuming to engineer manually. By automating the state preparation process, researchers can efficiently handle complex quantum states, unlocking new possibilities and accelerating progress in quantum information processing.


\item The final contribution of this work involves the application of approximation techniques, with the final goal of reducing the gate count of the generated quantum circuits. Specifically, these techniques target the elimination of operations that generate portions of the state that are practically irrelevant for the computation, as a result of the reduction in the size of the decision diagram representing the desired state. While previous work has focused on this aspect for \emph{qubit} circuit simulations~\cite{Hillmich2022}, the current study extends and generalizes these routines to encompass qudit systems, for a new purpose. 
Additionally, new reduction rules are introduced in the decision diagram simplification process.
The primary outcome of these approximation techniques is the simplification of the final logic. By eliminating nodes in the decision diagram, the quantum circuit complexity is also reduced, led by the discovery of patterns of tensor operations between sub-spaces of the single qudit Hilbert spaces, in the form of DD subgraphs. These intricate patterns are challenging to compute and have not been explored previously. Furthermore, a secondary benefit achieved is the reduction in the number of controls needed for each operation, enabling the translation to more resource-efficient sequences of operations.

\end{itemize}

\section{State Preparation}
\label{sec:implementation}

In this section, we propose the method to synthesize state decision diagrams into mixed-dimensional qudit circuits. 
To this end, we start by introducing decision diagrams, which efficiently store multi-dimensional quantum states. Then, we explain how these decision diagrams are used to create quantum circuits that generate the desired states. Lastly, we discuss how the resulting circuits can be further optimized using approximation techniques.

\subsection{Decision Diagrams}
\label{sec:decision-diagrams}
Previously, decision diagrams have been demonstrated to facilitate efficient representations of exponentially large data in numerous instances~\cite{DBLP:books/daglib/0027785,Miller2006,DBLP:journals/tcad/NiemannWMTD16,DBLP:conf/iccad/ZulehnerHW19}, owing to their ability to achieve compactness by capitalizing on the redundancy present in the data they depict. 

More precisely, a \emph{Decision Diagram} (DD) is a type of \emph{Directed Acyclic Graph} (DAG) comprising nodes and directed edges. 
These nodes are structured in levels, where each level corresponds to a specific qudit. 
The edges represent the choice of value for a qudit (or node) on a level and carry additional information, e.g.,~complex edge weights. The fundamental concept behind using decision diagrams to describe quantum states relies on recursively breaking down the associated vector.

\begin{figure}[tpb]
    \centering
    \includegraphics[width=0.8\linewidth]{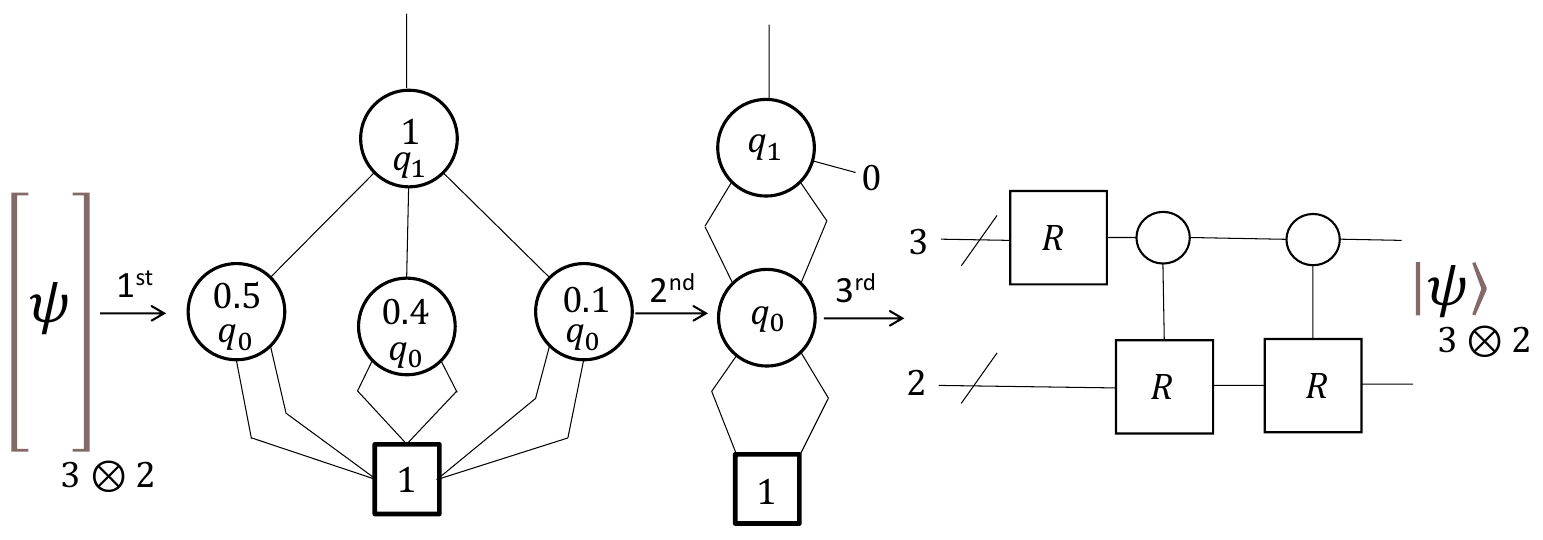}
    \caption{The three steps of state preparation.}
    \label{fig:contribution}
\end{figure}

This following procedure describes the first contribution and part of the procedure corresponding to the first arrow in \autoref{fig:contribution}, in other words, the representation of a quantum state as a decision diagram.
Consider an $n$-qudit quantum state defined over a set of variables $q_{n-1}, q_{n-2}, \ldots, q_0$, where $q_{n-1}$ w.l.o.g. is assigned as the \enquote{most sigificant qudit}.
Firstly, the state is divided into parts of equal size, based on the dimensionality of qudit $q_{n-1}$. Each part is associated with a successor node, which also contains the decision regarding the value of $q_{n-1}$. This splitting process continues for each sub-vector until the individual complex entries in the original vector are obtained. Throughout this process, the complex amplitudes of each basis state are stored in the form of weights inside the edges on the path from the root to the leaf followed during the decomposition. To ensure consistency, the nodes are normalized such that the sum of the squared magnitudes of the out-edges from a node adds up to one. To reconstruct the amplitude for specific basis states, one can traverse the decision diagram accordingly, multiplying all edge weights along the chosen path.

The resulting decision diagram consists of a \emph{root node} $q_{n-1}$, at least one node for each subsequent $q_i$, and ultimately a single \emph{terminal node} without any successors.
An example illustrates the procedure.

\begin{figure}[tpb]
    \centering
    \includegraphics[width=0.45\linewidth]{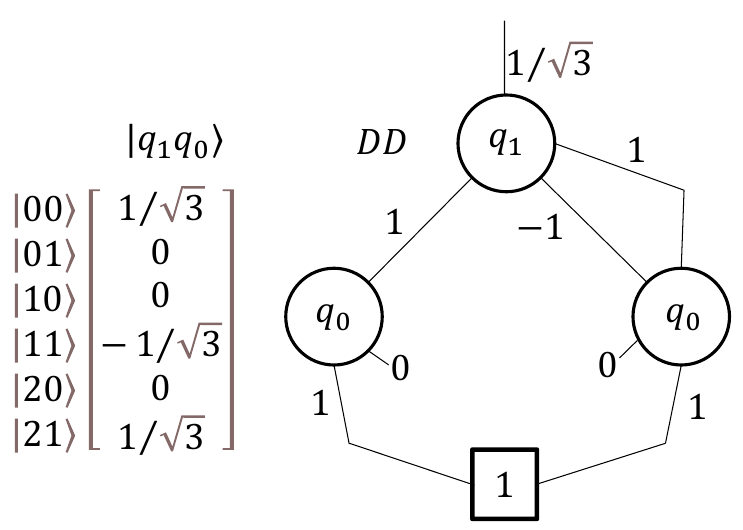}
    \caption{A state vector of circuit composed of a qutrit and a qubit and the corresponding decision diagram representation.}
    \label{fig:statedd}
\end{figure}
    
\begin{example}
    Consider the quantum state $\frac{1}{\sqrt{3}}(\ket{00} - \ket{11} + \ket{21})$ in a qutrit-qubit system, represented in 2 forms in \autoref{fig:statedd}.
    The vector's dimension is 6, which results from combining the local dimensionalities of the qutrit $3$ and the qubit $2$.
    The root node, denoted as $q_1$, branches into 3 edges, each corresponding to a distinct level in the qutrit.
    Moving to the $2^{nd}$ level, we encounter nodes representing the qubit, with each node having 2 edges.
    Notably, the 2nd and 3rd edges of the root node connect to the same qubit node, making use of redundancy.
    The nodes labeled as $q_0$ direct to the \emph{terminal node}.
    To calculate the amplitude of a specific basis state, we multiply the weights along the path corresponding to that basis state. For instance, for the bitstring $\ket{11}$, the computation involves multiplying $\nicefrac{1}{\sqrt{3}} \cdot -1 \cdot 1$, which accounts for the weights of the root node, the 1st edge of $q_1$, and the 1st edge of $q_0$.
\end{example}

\subsection{Synthesis Algorithm}
The algorithm initiates by recursively constructing the decision diagram that represents the quantum state, accounting for the dimensionalities of individual qudits and qubits, as described in the previous section. 
Due to varying dimensionalities among units of information, nodes at the same level may possess the same number of successors, while nodes at different levels could exhibit distinct dimensionality. 
Each connection linking a parent node and its successor is assigned a complex number weight--the normalizing factor calculated recursively from the out-edges of the successor. The normalization starts from the terminal nodes' edges. 
The weights are normalized by a fixed scheme to ensure canonicity in the nodes and subsequently a more compact representation after the reduction step. To normalize the weights of the out-edges of a given nodes, each weight is divided by the norm, such that the sum of squared magnitudes of the edge weights equals one. The norm is then multiplied to all weights on in-edges of the considered node.
At this point, the decision diagram forms a weighted tree and is fully prepared for the traversal, that will produce the quantum operations for constructing the desired quantum state.

Now we proceed by illustrating the generation of the quantum circuit starting from the decision diagram, contribution of this work corresponding to the third arrow in \autoref{fig:contribution}.
The routine operates recursively, starting from the root node, and traverses the decision diagram. It iterates through the successor edges, beginning from the end of the list, in pairs of two, following a decreasing order. At each step, the algorithm performs a two-level rotation, known as a Givens rotation\cite{ringbauer2021universal}, on the two adjacent edges or dimensions represented by the node.
The rotations between two levels $i$ and $j$ are expressed as 
\begin{align*}
    R_{i,j}(\theta,\phi) = \exp\left(\frac{-i \theta}{2} \Big(\cos(\phi)\sigma_x^{i,j} + \sin(\phi)\sigma_y^{i,j}\Big)\right),\label{eq:SUPPlocalOps}
\end{align*}
where $\sigma_{\{x,y\}}$ are the Pauli-$\{X,Y\}$ matrices, $\theta$ is the rotation angle, and $\phi$ is the phase of the rotation.
These parameters are calculated as:
$\theta = 2\cdot\arctan[|w_i/w_j|]$,
$\phi = -( \pi/2 + \arg[w_j] - \arg[w_i])$.
The sequence finishes with a phase rotation applied on the level 0-1, but with angle as the phase difference between the level 0, calculated from the sequence, and the father's node weight. The rotation can be decomposed into two-level rotations using the identity $\qop{Z}(\theta)= R(\tfrac{-\pi}{2}, 0)\cdot R(\theta,\tfrac{\pi}{2} )\cdot R(\tfrac{\pi}{2}, 0)$.

Every rotation is going to be applied to the circuit with a set of qudit controls equivalent to the set of nodes encountered on the path from the root to the node of the edges considered. For each node, the control level of the operation is going to be the index of the edge taken in order to descend the decision diagram.

\begin{figure}[tpb]
    \centering
    \includegraphics[width=0.5\linewidth]{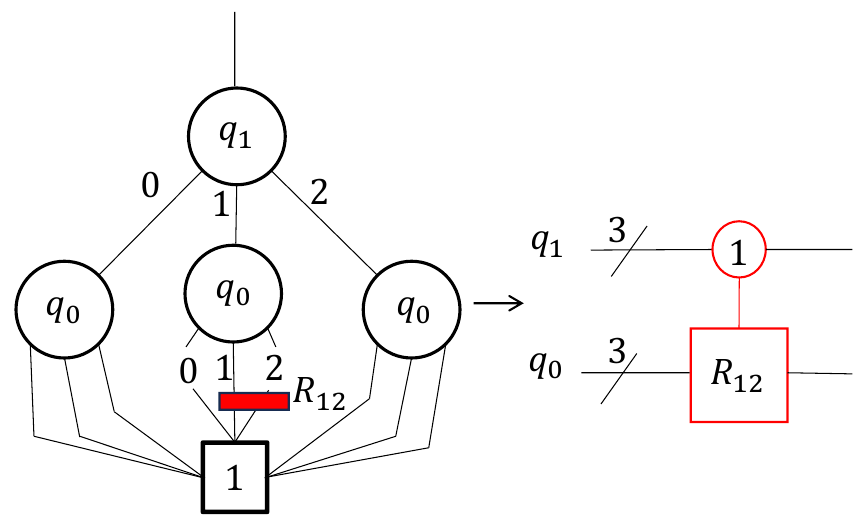}
    \caption{Example of a DD representing the state of a circuit composed of two qutrits and of a rotation synthesized from it.}
    \label{fig:controlled}
    \vspace{-1em}
\end{figure}

\begin{example}
    \autoref{fig:controlled} shows a step of the synthesis algorithm. The algorithm appends to the circuit the rotation R on levels 1 and 2, and calculates the parameters accordingly. The rotation is controlled on level 1 of the the first qutrit since the rotation was derive from the node with index 1.
\end{example}

\subsection{Optimization and Approximation}
In this section we present part of the contribution of this work that is depicted by the second arrow in \autoref{fig:contribution}.
To facilitate the synthesis of shorter circuits, we employed several optimizations during the decision diagram construction stage. 
The first optimization comes from studies in quantum circuit simulation.
This field has exploited decision diagrams in the past and it requires efficient approaches due to exponential complexity of the problem tackled. 
Approximated decision diagrams proved to be an effective solution for improving memory complexity and runtime performance~\cite{Hillmich2022}. 
However, the technique comes with a trade-off with state fidelity, requiring a careful balance between computational efficiency and accuracy in results. 
The technique is a generalization of~\cite{Hillmich2022}, where it calculates the contributions in terms of fidelity of each node and then removing nodes from the decision diagram until a threshold fidelity, previously chosen for the synthesis, is reached. The contribution is calculated as the sum of the squared magnitude of each amplitude that has a path from root to leaves crossing the node.

The benefits of the technique are the following:
\begin{itemize}
    \item The method reduces the size of the decision diagram in terms of memory.
    \item It reduces the synthesis time, since its complexity is tied to the size of the decision diagram.
    \item Lastly, the technique enables synthesizing shorter circuits for the state preparation, with guarantees on the fidelity reached by the procedure
\end{itemize}

The second optimization consists in the reduction of the decision diagram, intended as the capability of two edges pointing to the same node, whenever it represents two identical sub-trees, that would be otherwise stored twice. The advantage here lies in the logical significance of this reduction. When all edges with weights other than zero point to the same node, it resembles a tensor product operation between the two qudits representing adjacent levels in the tree. Consequently, operations in the sub-tree will not consider the father node with successors pointing to the same child as a control qudit, thereby reducing the number of entangling gates during transpilation.

\begin{example}
    \autoref{fig:contribution} depicts the application of the optimization techniques. 
    The successor node with lowest fidelity contribution (0.1) of the root is pruned from the decision diagram.
    The two remaining edges that point to nodes representing the same sub-vectors are redirected to the same shared node. 
    This will reduce the amount of memory used for storing nodes, and due to the properties of tensor products, no controls will be synthesized.
\end{example}

\medskip
Overall, the method results in a state preparation procedure of three main steps, as depicted in \autoref{fig:contribution}: starting with a representation of the state as a decision diagram, followed by an approximation routine to reduce the size of the DD, and concluding with the realization of optimized quantum circuits.

\section{Experimental Evaluation}
\label{sec:evaluation}

\begin{table*}[tbp]
    \centering
    \caption{Evaluation of the proposed approach comparing the average results over 40 runs of the synthesis method per benchmark}
    \vspace{-0.5em}
    \label{tbl:results}
    \resizebox{\linewidth}{!}{%
    \begin{tabular}{@{}lrl@{\hspace{1em}}rrrrr@{\hspace{1.5em}}rrrrrr@{}}
        \toprule
        \multicolumn{3}{c}{Benchmark} & \multicolumn{5}{c}{Exact (Averaged)} & \multicolumn{6}{c}{Approximated 98\% (Averaged)} \\
         \cmidrule(r{1em}){1-3}\cmidrule(r{1em}){4-8}\cmidrule{9-14}
    	Name & \#Qudits & {Qudits}  & Nodes & Distinct$\mathbb{C}$ & Operations &  \#Controls  & Time [s] & Nodes & Distinct$\mathbb{C}$ & Operations &  \#Controls  & Time [s] & Fidelity\\ \midrule

    Emb. W-State&3&$[1 \times 3, 1 \times 6, 1 \times 2]$&58.0&5.0&21.0&2.0&0.00&22.0&6.0&21.0&2.0&0.00&1.00\\
                &4&$[1 \times 9, 1 \times 5, 1 \times 6, 1 \times 3]$&1135.0&7.0&49.0&3.0&0.01&50.0&8.0&49.0&3.0&0.01&1.00\\
                &6&$[3 \times 4, 1 \times 7, 1 \times 3, 1 \times 5]$&8657.0&12.0&91.0&3.0&0.03&92.0&12.0&91.0&3.0&0.03&1.00\\[4pt]
    
    GHZ State&3&$[1 \times 3, 1 \times 6, 1 \times 2]$&58.0&3.0&19.0&2.0&0.00&20.0&3.0&19.0&2.0&0.00&1.00\\
            &4&$[1 \times 9, 1 \times 5, 1 \times 6, 1 \times 3]$&1135.0&3.0&51.0&2.0&0.01&52.0&3.0&51.0&2.0&0.01&1.00\\
            &6&$[3 \times 4, 1 \times 7, 1 \times 3, 1 \times 5]$&8657.0&3.0&73.0&2.0&0.04&74.0&3.0&73.0&2.0&0.05&1.00\\[4pt]
    
    W-State&3&$[1 \times 3, 1 \times 6, 1 \times 2]$&58.0&5.0&37.0&2.0&0.00&38.0&6.0&37.0&2.0&0.00&1.00\\
            &4&$[1 \times 9, 1 \times 5, 1 \times 6, 1 \times 3]$&1135.0&11.0&186.0&2.0&0.03&185.0&10.0&186.0&2.0&0.03&1.00\\
            &6&$[3 \times 4, 1 \times 7, 1 \times 3, 1 \times 5]$&8657.0&14.0&262.0&4.0&0.06&259.0&14.0&262.0&4.0&0.06&1.00\\[4pt]
    
      Random State&3&$[1 \times 3, 1 \times 6, 1 \times 2]$&58.0&58.0&57.0&2.0&0.00&40.2&52.9&54.05&1.9&0.00&0.99\\
    &4&$[1 \times 9, 1 \times 5, 1 \times 6, 1 \times 3]$&1135.0&1135.0&1134.0&2.0&0.12&573.92&1071.97&1084.28&2.82&0.10&0.99\\
    &5&$[2 \times 6, 1 \times 5, 2 \times 3]$&2383.0&2383.0&2382.0&4.0&0.14&1255.22&2276.62&2287.07&3.75&0.16&0.99\\
    &6&$[3 \times 5, 1 \times 4, 2 \times 2]$&3266.0&3266.0&3265.0&5.0&0.18&2187.15&3121.65&3136.9&4.79&0.19&0.99\\
    &6&$[3 \times 4, 1 \times 7, 1 \times 3, 1 \times 5]$&8657.0&8657.0&8656.0&5.0&0.44&3176.65&8336.8&8357.35&4.78&0.39&0.99\\
    
         \bottomrule
    \end{tabular}
    }
    \vspace{-1em}
\end{table*}

To evaluate the feasibility of the proposed quantum state preparation approach, we implemented the method and evaluated it on several types of quantum states.
The selected benchmarks are
\begin{itemize}
    \item Embedded W-State~\cite{yeh2023scaling},
    \item GHZ State~\cite{greenberger2007going},
    \item W-State~\cite{Cabello2002}, and
    \item Random states with amplitudes generated from a uniform distribution.
\end{itemize}
For each state a mixed-dimensional system is used as target architecture.
The implementation is written in Python~3, excluding third-party dependencies.
The evaluations were performed on a server running GNU/Linux using an AMD Ryzen Threadripper PRO 5955WX (at \SI{4}{\giga\hertz}) and \SI{125}{\gibi\byte}~main~memory. 
The implementation used in this evaluation is freely available (as part of the Munich Quantum Toolkit, MQT) at \href{https://github.com/cda-tum/mqt-qudits}{\emph{github.com/cda-tum/mqt-qudits}}. 

\smallskip

The results are presented in \autoref{tbl:results}.
The synthesis procedure is executed \num{40} times to average possible fluctuations. 
The initial group of columns provides key information about the benchmark, including the algorithm's name, the number of qudits involved, and the qudit dimensions represented as \mbox{\emph{Count $\times$ Dimension}}, randomly selected to further prove the efficacy of the method. For example, an entry $2\times4$ signifies two qudits, each with a dimension of four.
The following two column groups present details on the resulting qudit circuits for both \enquote{Exact} (circuits that compile the target state with fidelity~$1$) and \enquote{Approximated \SI{98}{\percent}} (compiling the target state with fidelity of at least $0.98$). 
The columns \enquote{Nodes} and \enquote{Distinct$\mathbb{C}$} represent the average number of nodes and unique complex numbers in the decision diagram representing the state.
These are the main metrics for evaluating the efficiency in the decision diagram for storing the \mbox{mixed-dimensional} quantum states. 
The \enquote{Operations} column indicates the average number of multi-controlled operations synthesized for a state on a specific mixed-dimensional architecture, relevant for both exact and approximated synthesis. 
Additionally, the column \enquote{Controls} refers to the average median number of controls present in multi-controlled operations synthesized for a state on the mixed-dimensional architecture. 
The use of controlled operations as a primary metric is justified by the fact that the circuit can later be transposed into a sequence of local and two-qudit operations~\cite{Di2013}, with also linear complexity in terms of depth, as demonstrated in \cite{zi2023optimal}.
Finally, the \enquote{Time} column denotes the elapsed time during the approximation and synthesis process. 
The method is efficient, since the synthesis routine has time complexity linear in the number of nodes of the DD. Moreover, the approximation technique leads to a linear improvement in the number of quantum gates, with the reduction of control nodes during the synthesis. 

\smallskip

Both approaches, \enquote{Exact} and \enquote{Approximated} have been shown to be capable of preparing a given quantum state in a mixed-dimensional system.
Due to the regular structure of the first three (non-random) benchmarks, the approximation shows no effect.
In contrast, approximation decreases the number of operations (and controls) by about \SI{5}{\percent} while losing only \SI{1}{\percent} fidelity.
Regardless of the approximation step, the individual runs of the benchmarks finish in less than one second.

\smallskip

Overall, the results confirm the successful development of the first automated method for synthesizing quantum circuits that can construct arbitrary states for \mbox{mixed-dimensional} quantum systems. Furthermore, the outcomes demonstrate the method's efficiency and the capabilities in optimizing the circuits, preparing the desired state without significantly affecting its fidelity.

\section{Conclusion}
\label{sec:conclusion}
In this work, we proposed a novel synthesis method for the generation of quantum circuits, performing the state preparation of arbitrary states for \mbox{mixed-dimensional} quantum systems.  
This is the first developed automated method for such an application on quantum architectures with qubits and qudits of different sizes.

The approach uses edge-weighted decision diagrams with a variable number of successors for the representation of the state and for the synthesis of the quantum circuits. Extended approximation techniques applied to the data-structure optimizes the number of operations in the circuits.
The evaluations conducted in the study showcased the method's efficiency and versatility in preparing arbitrary quantum states for any \mbox{mixed-dimensional} quantum architecture. 
Through optimizations, we achieved a notable reduction in the number of operations and controls required, while maintaining a high fidelity.
Potential future work could involve further refining and exploring qudit quantum circuit optimization and approximation techniques by taking the capabilities of the targeted quantum hardware in account.


\section*{Acknowledgment}
{\small This work has received funding from the European Union’s Horizon 2020 research and innovation programme under the ERC Consolidator Grant (agreement No~101001318) and the NeQST Grant (agreement No~101080086).
It is part of the Munich Quantum Valley, which is supported by the Bavarian state government with funds from the Hightech Agenda Bayern Plus and was partially supported by the BMK, BMDW, and the State of Upper Austria in the frame of the COMET program (managed by the FFG).}

\printbibliography
\end{document}